\documentclass[aps,prb,reprint,groupedaddress]{revtex4-1}

\usepackage[english]{babel}
\usepackage{subfiles}

\usepackage{amsmath}
\newcommand{\angstrom}{\textup{\AA}}

\usepackage{graphicx}
\usepackage{bm}
\usepackage{amssymb, amsmath}
\usepackage{algorithm}
\usepackage{algpseudocode}
\usepackage{url}

\begin{document}


\title{A local Bayesian optimizer for atomic structures}


\author{Estefan\'ia Garijo del R\'io}
\author{Jens J{\o}rgen Mortensen}
\author{Karsten Wedel Jacobsen}
\affiliation{CAMD, Department of Physics, Technical University of Denmark}


\date{\today}

\begin{abstract}
  A local optimization method based on Bayesian Gaussian processes is
  developed and applied to atomic structures. The method is applied to
  a variety of systems including molecules, clusters, bulk materials,
  and molecules at surfaces. The approach is seen to compare favorably
  to standard optimization algorithms like conjugate gradient or BFGS
  in all cases. The method relies on prediction of surrogate potential
  energy surfaces, which are fast to optimize, and which are gradually
  improved as the calculation proceeds. The method includes a few
  hyperparameters, the optimization of which may lead to further
  improvements of the computational speed.
\end{abstract}

\pacs{}

\maketitle

\section{Introduction}
One of the great successes of density functional theory (DFT)
\cite{Hohenberg:1964fz,Kohn:1965js} is its ability to predict ground
state atomic structures.  By minimizing the total energy, the atomic
positions in solids or molecules at low temperatures can be
obtained. However, the optimization of atomic structures with density
functional theory or higher level quantum chemistry methods require
substantial computer resources. It is therefore important to develop
new methods to perform the optimization efficiently.

It is of key interest here, that for a given atomic structure a DFT
calculation provides not only the total electronic energy, but also,
at almost no additional computational cost, the forces on the atoms,
\emph{i.e.}\ the derivatives of the energy with respect to the atomic
coordinates. This means that for a system with $N$ atoms in a
particular configuration only a single energy-value is obtained while
$3N$ derivatives are also calculated. It is therefore essential to
include the gradient information in an efficient optimization.

A number of well-known function optimizers exploring gradient
information exist \cite{Press:2007te} and several are implemented in
standard libraries like the SciPy library \cite{scipy} for use in
Python. Two much-used examples are the conjugate gradient (CG) method
and the Broyden--Fletcher--Goldfarb--Shanno (BFGS) algorithm. Both of
these rely on line minimizations and perform particularly well for a
nearly harmonic potential energy surface (PES). In the CG method, a
series of conjugated search directions are calculated, while the BFGS
method gradually builds up information about the Hessian, \emph{i.e.}\
the second derivatives of the energy, to find appropriate search
directions.

The Gaussian process (GP) method that we are going to present has the
benefit that it produces smooth surrogate potential energy surfaces
(SPES) even in regions of space where the potential is
non-harmonic. This leads to a generally improved convergence.  The
number of algebraic operations that has to be carried out in order to
move from one atomic structure to the next is much higher for the GP
method than for the CG or BFGS methods, however, this is not of
concern for optimizing atomic structures with DFT, because the
electronic structure calculations themselves are so time
consuming. For more general optimization problems where the function
evaluations are fast, the situation may be different.

Machine learning for PES modelling has recently attracted the
attention of the materials modelling community \cite{Bartok:2017hz,
  Huang:2017ub, Glielmo:2018bm, Behler:2007fe, Rupp:2012kx,
  Csanyi:2004dh, Khorshidi:2016fu, BO-rrss-2018, hannesGP,
  Hammer-evolutionary-2018, Rinke-preprint-2017, Karsten-Bjork-2018, Podryabinkin:2017jp, shapeev:alloys}.
In particular, several methods have focused on fitting the energies
of electronic structure calculations to expressions of the form
\begin{equation}
E(\rho) = \sum_{i=1}^n \alpha_i\, k\left(\rho^{(i)},\; \rho\right).
\label{eq:intro_to_krr}
\end{equation}

Here, $\{\rho^{(i)}\}_{i=1}^{n}$ are some descriptors of the $n$
atomic configurations sampled, $k\left(\rho^{(i)},\; \rho\right)$ is
known as a kernel function and $\{\alpha_i\}_{i=1}^n$ are the
coefficients to be determined in the fit.  Since there are $n$
coefficients and $n$ free parameters, the SPES determined by this
expression has the values of the 
calculations at the
configurations on the training set.

Here we note that expression \eqref{eq:intro_to_krr} can easily be
extended to:
\begin{equation}
  E(\rho) = \sum_{i=1}^n \alpha_i\, k\left(\rho^{(i)},\; \rho\right) + \sum_{i=1}^n \sum_{j=1}^{3N} \beta_{ij} \; \frac{\partial k\left(\rho^{(i)},\; \rho\right)}{\partial r_j^{(i)}} , 
\label{eq:Thorsten}
\end{equation}
where $\{r_j^{(i)}\}_{j=1}^{3N}$ represent the coordinates of the N atoms
in the $i-$th configuration.
The new set of parameters $\beta_{ij}$ together with $\alpha_i$
can be adjusted so that not only the right energy of a given
configuration $\rho^{(i)}$ is predicted, but also the right
forces. This approach has two advantages with respect to the previous
one: (i) the information included in the model scales with the
dimensionality, (ii) the new model is smooth and has the right
gradients.

In the case of systems with many identical atoms or similar local
atomic structures it becomes advantageous to construct SPESs based on
descriptors or fingerprints characterizing the local environment
\cite{Bartok:2017hz, Huang:2017ub, Glielmo:2018bm, Behler:2007fe,
  Rupp:2012kx, Csanyi:2004dh, Khorshidi:2016fu}. The descriptors can
then be constructed to obey basic principles as rotational and
translational symmetries and invariance under exchange of identical
atoms. Here we shall develop an approach based on Gaussian processes which
works directly with the atomic coordinates and effectively produces a
surrogate PES of the type Eq.~\eqref{eq:Thorsten} aimed at relaxing
atomic structures. We note, that Gaussian processes with derivatives for PES modeling is a field that is developing fast, with recent applications in local optimization \cite{denzel2018gaussian} and path determination in elastic band calculations \cite{hannesGP, hannes2019NEB, Jose2019GP}.

\section{Gaussian process regression}
We use Gaussian process regression with derivative information to produce a
combined model for the energy $E$ and the forces $\mathbf{f}$ of a
configuration with atomic positions
$\mathbf{x} = ( \mathbf{r}_1, \mathbf{r}_2, \dots, \mathbf{r}_N)$:
\begin{equation}
  \mathbf{U}(\mathbf{x}) = \left(E(\mathbf{x}), -\mathbf{f}(\mathbf{x})\right) \sim 
  \mathcal{GP}\left(\mathbf{U}_p(\mathbf{x}), K(\mathbf{x},\mathbf{x}^\prime)\right),
\end{equation} 
where
$\mathbf{U}_p(\mathbf{x}) = (E_p(\mathbf{x}), \nabla E_p(\mathbf{x}))$
is a vector-valued function which constitutes the prior model for the
PES and $K(\mathbf{x},\mathbf{x}^\prime)$ is a matrix-valued kernel
function that models the correlation between pairs of energy and force
values as a function of the configuration space.

In this work, we choose the constant function
$\mathbf{U}_p(\mathbf{x}) = (E_p, \mathbf{0})$ as the prior
function. For the kernel, we use the squared-exponential covariance
function to model the correlation between the energy of different
configurations:
\begin{equation}
  k(\mathbf{x}, \mathbf{x}^\prime) = 
  \sigma^2_f e^{-\Vert \mathbf{x} -\mathbf{x}^\prime \Vert ^2/2l^2},
  \label{eq:squared-exponential}
\end{equation}
where $l$ is a typical scale of the problem and $\sigma_f$ is a
parameter describing the prior variance at any configuration
$\mathbf{x}$. The full kernel $K$ can be obtained by noting that
\cite{rasmussen2006a, NIPS2017_7111}:
\begin{align}
  \mathrm{cov} \left(E(\mathbf{x}), E(\mathbf{x}^\prime)\right) &= k(\mathbf{x}, \mathbf{x}^\prime) &\\
  \mathrm{cov} \left(E(\mathbf{x}), \frac{\partial E(\mathbf{x}^\prime)}{\partial x^\prime_i} \right) &= \frac{\partial k(\mathbf{x}, \mathbf{x}^\prime)}{\partial x^\prime_i} & \equiv J_i(\mathbf{x},\mathbf{x}^\prime)\\
  \mathrm{cov}  \left(\frac{\partial E(\mathbf{x})}{\partial x_i}, \frac{\partial E(\mathbf{x}^\prime)}{\partial x^\prime_j} \right) &= \frac{\partial^2 k(\mathbf{x}, \mathbf{x}^\prime)}{\partial x_i\partial x^\prime_j} & \equiv H_{ij}(\mathbf{x},\mathbf{x}^\prime),
\end{align}
and assembling these covariance functions in a matrix form: 
\begin{equation}
K(\mathbf{x}, \mathbf{x}^\prime) = \left( \begin{array}{cc}
k(\mathbf{x} , \mathbf{x}^\prime)          &  \mathbf{J}(\mathbf{x} , \mathbf{x}^\prime)\\
\mathbf{J}(\mathbf{x}^\prime , \mathbf{x})^T &  H(\mathbf{x} , \mathbf{x}^\prime)
\end{array} \right).
\end{equation}

The expressions for the mean and the variance for the posterior
distribution follow the usual definitions incorporating the additional
matrix structure. Let $X = \{ \mathbf{x}^{(i)}\}_{i=1}^{n}$ denote the
matrix containing $n$ training inputs and let
$Y=\{\mathbf{y}^{(i)}\}_{i=1}^n =\{ \left(E(\mathbf{x}^{(i)}),
  -\mathbf{f}(\mathbf{x}^{(i)})\right)\}_{i=1}^n $ be the matrix
containing the corresponding training targets.  By defining
\begin{equation}
K(\mathbf{x}, X) = \left( K(\mathbf{x}, \mathbf{x}^{(1)}),  K(\mathbf{x}, \mathbf{x}^{(2)}), \dots,  K(\mathbf{x}, \mathbf{x}^{(n)})\right)
\end{equation}
and
\begin{equation}
\left(K(X,X)\right)_{ij} = K(\mathbf{x}^{(i)}, \mathbf{x}^{(j)}),
\end{equation}
we get the following expressions for the mean:
\begin{align}
\bar{\mathbf{U}}(\mathbf{x}) &=
(\bar{E}(\mathbf{x}), -\bar{\mathbf{f}}(\mathbf{x}))\label{eq:prediction}\nonumber \\
&= \mathbf{U}_p(\mathbf{x}) + K(\mathbf{x}, X) \mathbb{K}_X^{-1}(Y - \mathbf{U}_p(X))
\end{align}
and the variance:
\begin{equation}
\bm{\sigma}^2(\mathbf{x}) = K(\mathbf{x}, \mathbf{x}) - K(\mathbf{x}, X)\mathbb{K}_X^{-1}K(X,\mathbf{x})
\label{eq:uncertainty}
\end{equation}
of the prediction, where $\mathbb{K}_X = K(X,X)+\Sigma_n^2$. Here, we have assumed an additive Gaussian noise
term with covariance matrix $\Sigma_n$ \cite{rasmussen2006a}. 
This term corrects only for the self covariance of the points in the training set, and thus,
it is a diagonal matrix that models the self correlation of forces 
with a hyperparameter  $\sigma_n^2$ and the self correlation of energies with 
$\sigma_n^2\times l^2$.  We note that even
for computational frameworks where the energy and forces can be
computed with very limited numerical noise, small non-zero values of
$\sigma_n$ are advantageous since they prevent the inversion of the
covariance matrix $K(X,X)$ to be numerically ill-conditioned
\cite{hannesGP}.

In the following, we will refer to $\bar{E}(\mathbf{x})$ as defined in equation \eqref{eq:prediction} as the surrogate potential energy surface (SPES) and distinguish it from the first principles PES, $E(\mathbf{x})$.

\section{Gaussian process minimizer: GPMin}
The GP framework can be used to build an optimization algorithm. In this section, 
we introduce the main ideas behind the proposed Gaussian process minimizer (denoted GPMin from  hereon). A more detailed description of the algorithm
can be found in the Appendix in the form of pseudocode.

The GP regression provides a SPES that can be minimized
using a gradient-based local optimizer. For this purpose, we have used
the L-BFGS-B algorithm as implemented in SciPy \cite{lbfgsb}. The
prior value for the energy is initially set as the energy of the
initial configuration and then the expression \eqref{eq:prediction} is
used to produce a SPES from that data point alone. This model is then
minimized, and the evaluation at the new local minimum generates new
data that is then fed into the model to produce a new SPES that will
have a different local minimum. Before generating each new SPES the
prior value for the energy is updated to the maximum value of the
energies previously sampled. This step is important because it makes the algorithm more stable. If a high-energy configuration is sampled, the forces may be very large leading to a too large new step. The increase of the prior value tends to dampen this by effectively reducing the step size. The whole process is then iterated until convergence is reached. 

It is illustrative to consider in more detail the first step of the algorithm. It is straightforward to show using equations~\eqref{eq:squared-exponential} - \eqref{eq:prediction} that if only a single data point $\mathbf{x}^{(1)}$ is known the SPES is given by
\begin{equation} \label{eq:one-point-E}
    \bar{E}(\mathbf{x}) = E^{(1)} -\mathbf{f}^{(1)}\cdot(\mathbf{x} - \mathbf{x}^{(1)}) \,e^{-\Vert \mathbf{x} - \mathbf{x}^{(1)}\Vert^2/2l^2},
\end{equation}
where $E^{(1)}$ and $\mathbf{f}^{(1)}$ are the energy and forces of the SPES at the point $\mathbf{x}^{(1)}$, respectively. We have here used that the prior energy is set to the energy of the first configuration $E^{(1)}$. One can confirm that this is the prior energy by noting that points far away from $\mathbf{x}^{(1)}$, where no information is available, take on this value for the energy. It is seen that the initial force $\mathbf{f}^{(1)}$ gives rise to a Gaussian depletion of the SPES. The first step of the GPMin algorithm minimizes the SPES leading to a new configuration
\begin{equation} \label{eq:stat_points}
    \mathbf{x} =  \mathbf{x}^{(1)} + l \frac{\mathbf{f}^{(1)}}{\Vert\mathbf{f}^{(1)}\Vert}.
\end{equation}
The first step is thus in the direction of the force with a step length of $l$. Considering the information available this is a very natural choice.

GPMin depends on a number of parameters: the length scale $l$, the prior value of the energy $E_p$, the energy width $\sigma_f$, and the noise or regularization parameter $\sigma_n$. 
It can be seen from expressions \eqref{eq:squared-exponential} and
\eqref{eq:prediction} that the prediction of the SPES depends only on the ratio of $\sigma_f$ and $\sigma_n$ and not their individual values.

The prior energy $E_p$ is, as explained above, taken initially as the energy of the first configuration and then updated if larger energies are encountered. It is important that the prior value is not too low to avoid large steps, since the prior energy is the value of the SPES for all configurations far away (on the scale of $l$) from previously investigated structures.

The scale $l$ is very important as it sets the distance over which the SPES relaxes back to the prior value $E_p$ when moving away from the region of already explored configurations. It therefore also effectively determines a step length in the algorithm. 

One interesting advantage of the Bayesian approach is that it allows for update of parameters (usually termed hyperparameters) based on existing data. We investigate this option by allowing the value of the length scale $l$ to change. Since the update procedure also depends on the width parameter $\sigma_f$, we update this as well. The updated hyperparameters,
$\bm{\theta} = (l, \sigma_f)$,
are determined by maximizing the marginal likelihood:
\begin{equation}
\bm{\theta} = \arg \max_{\bm{\vartheta}} P\left(Y\vert X,\bm{\vartheta}\right).
\label{eq:update}
\end{equation}

 The optimization may fail, for example if there is not enough evidence and the marginal likelihood is very flat, and if that happens, the previous scale is kept. The update procedure allows the algorithm to find its own scale as it collects more information, producing a model that self-adapts to the problem at hand.
In section \ref{sec:hyperparams}  we shall consider in more depth the adequate choices for the values of the hyperparameters and the different strategies for the update of hyperparameters when the optimizers are applied to DFT calculations.

\section{Computational details}
We illustrate and test the method on a variety of different systems
using two different calculation methods: An interatomic effective medium theory potential (EMT) \cite{Jacobsen:1987gh,Jacobsen:1996kb}
as implemented in ASE \cite{ase, ase-paper} and DFT.  The DFT tests
have been performed using GPAW \cite{gpaw} with the local density
approximation (LDA) exchange-correlation functional and a plane wave
basis set with an energy cutoff at 340 eV. The
Brillouin zone has been sampled using the Monkhorst-Pack scheme with a
k-point density of 2.0/$(\angstrom^{-1})$ in all three directions. The
PAW setup with one valence electron has been used for the sodium
cluster for simplicity. In addition to the default convergence criteria for GPAW, we specify
 that the maximum change in magnitude 
of the difference in force for each atom should be smaller than $10^{-4}
\textrm{eV}\angstrom^{-1}$ for the self-consistent field iteration to terminate. This improves the convergence of the forces.
All systems have been relaxed until the
maximum force of the atoms was below 0.01 eV$\angstrom^{-1}$.

\section{Example: Gold clusters described in effective medium theory}

In the first example GPMin is
used to find the structure of 10-atom gold clusters as described by
the EMT potential, and the efficiency is compared with other common
optimizers.  For this purpose, we generate 1000 random
configurations of a 10-atom gold cluster. The configurations are constructed by sequentially applying three uniform displacements for
each atom in a cubic box with side length 4.8$\angstrom$ and only keeping
those that lie further than 1.7 times the atomic radius of gold away
from any of the other atoms already present in the cluster.  Each
configuration is then optimized with different choices of parameters
for GPMin, and, for comparison, the same structures are optimized with
the ASE implementations of FIRE \cite{FIRE} and BFGS Line Search, and
the SciPy implementations of BFGS and the CG.

\begin{figure}
\centering
\includegraphics[scale=1]{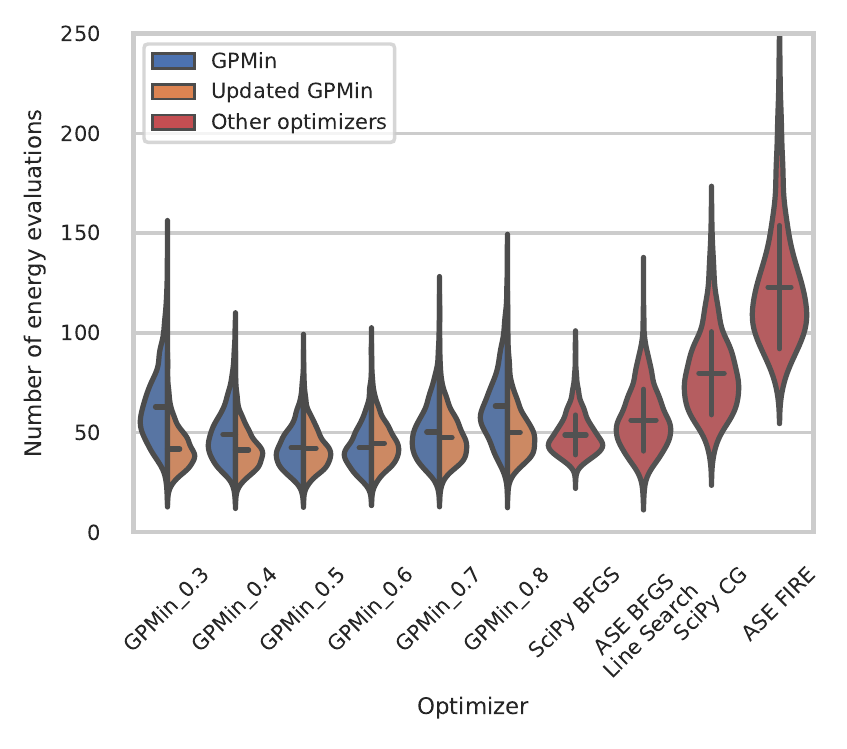}
\caption{Statistics of the number of energy evaluations 
for 1000 relaxations of a 10-atom gold cluster. The initial conditions have been randomly 
generated. The left hand side of the plot shows the distribution of the number of energy evaluations for GPMin in its two variants for scales ranging from 0.3 to 0.8 \angstrom: keeping the scale fixed or allowing it to be updated. The right hand side shows the performance of other widely used optimizers, which have been sorted according to the average number of function evaluations. 
\label{fig:violin}}
\end{figure}

For the gold clusters, we have investigated the effect of updating $\sigma_f$ and $l$ for six
different initial scales between $0.3$ and $0.8\,\angstrom$ and initial $\sigma_f=1.0$ eV. 
Since the EMT potential has very small numerical noise, we choose a small value of $\sigma_n/\sigma_f = 5\times 10^{-4} \textrm{eV}\angstrom^{-1}$ for the regularization. In the update-version of the optimizer, we update the scale every 5th
iteration.

The statistics of the number of energy evaluations are shown in
Figure~\ref{fig:violin}. The GP optimizers are seen to be the fastest
on average, with the appropriate choice of the hyperparameters. For the initial scale of 0.5$\angstrom$, for example, the updated version of GPMin had relaxed the clusters after  $42.1 \pm 0.3$ energy evaluations
and the non-updated one after  $42.5 \pm 0.3$, as compared to
$48.8 \pm 0.3$ and $56.2 \pm 0.5$ for the BFGS implementations in
SciPy and ASE, respectively. CG exhibits $79.7\pm 0.7$ average number
of steps and FIRE, $122.9\pm 1.0$.  

Figure \ref{fig:violin} shows the trend in the performance as the scale is varied. For this system, $l=0.5\angstrom$ has the lowest average and variance for GPMin. The performance depends rather sensitively on the scale parameter: reducing the scale results in a more conservative algorithm where more but smaller steps are needed. Increasing the scale leads to a more explorative algorithm with longer steps that may fail to reduce the energy.  In the algorithm with updates, the scale is automatically modified to compensate for a non-optimal initial scale. The update is particularly efficient for small scales where the local environment is sufficiently explored. For larger scales the sampling is less informative and it takes longer for the algorithm to reduce the scale. 

We note that under the appropriate choice of scale, both GPMin with and without update are  among the fastest for the best case scenario, with 18 evaluations for the regular GPMin optimizer and 19 for the updated version with scale $l=0.5$\angstrom, compared to 19
for ASE BFGS, 27 and 34 for the SciPy implementations of BFGS and CG
respectively and 70 for FIRE.  We further note that the updated
version has by far the best worst-case performance.  

Of the total of 18000 relaxations, only 17 failed to find a local minimum. These 17 relaxations were all run with the GPMin optimizer with $l=0.8\angstrom$ without the updates. An optimizer with a too long scale fails to build a successful SPES: the minimum of the SPES often has a higher energy than the previously evaluated point. Thus, we consider the optimization has failed if after 30 such catastrophic attempts, the optimizer has still not being able to identify a point that reduces the energy or if SciPy's BFGS cannot successfully optimize the predicted SPES.

\section{Determination of the hyperparameters} \label{sec:hyperparams}

\begin{figure}
\centering
\includegraphics[scale=0.65]{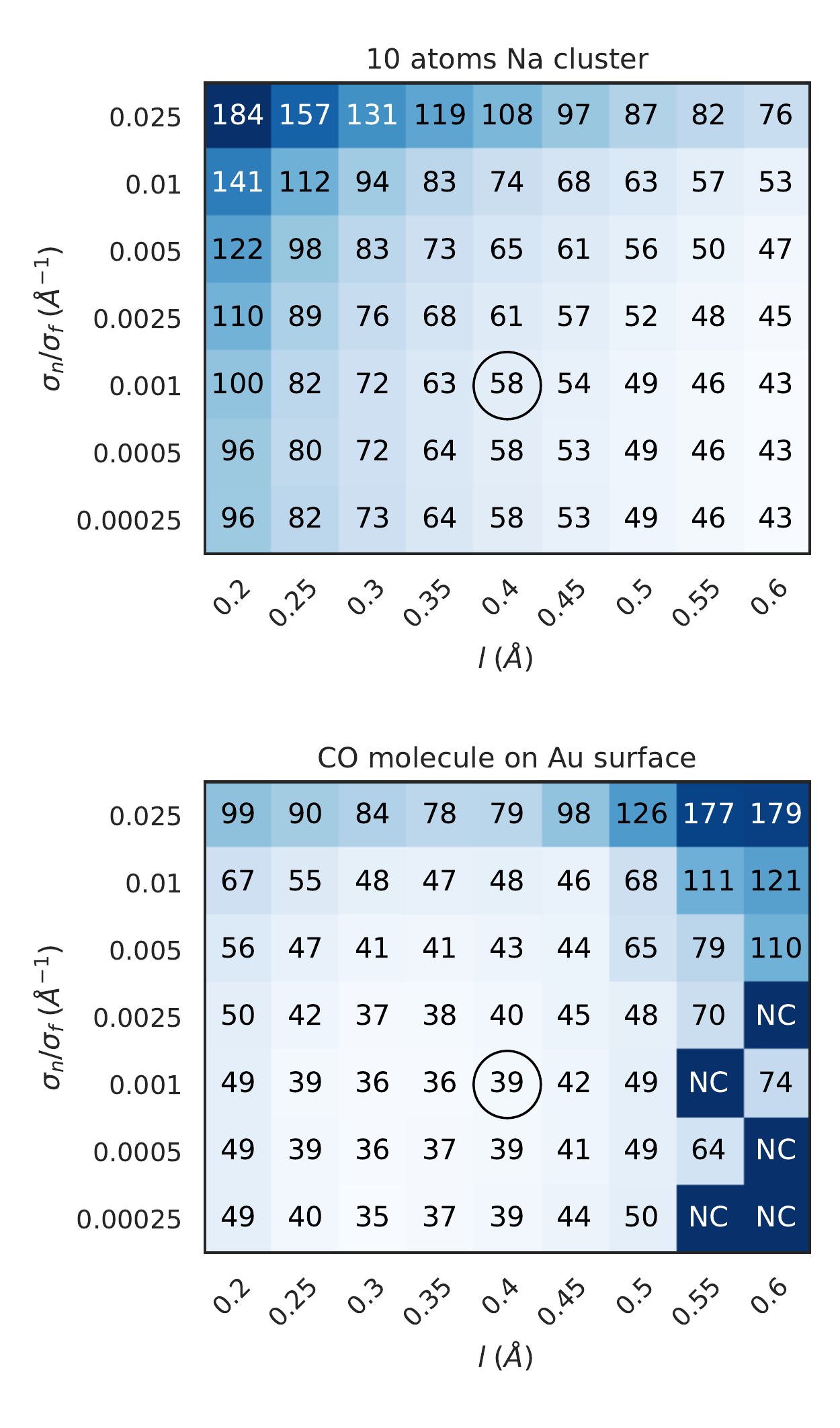}
\caption{Average number of potential energy evaluations needed to relax 10
  atomic structures as a function of the two hyperparameters: the
  length scale $l$, and the regularization parameter $\sigma_n$. The
  label NC (Not Converged) indicates that at least one of the 
  relaxations did not
  converge. The default choices for the hyperparameters are indicated
  by circles.\label{fig:params}}
\end{figure}

\begin{figure*}
    \centering
    \includegraphics[scale=1]{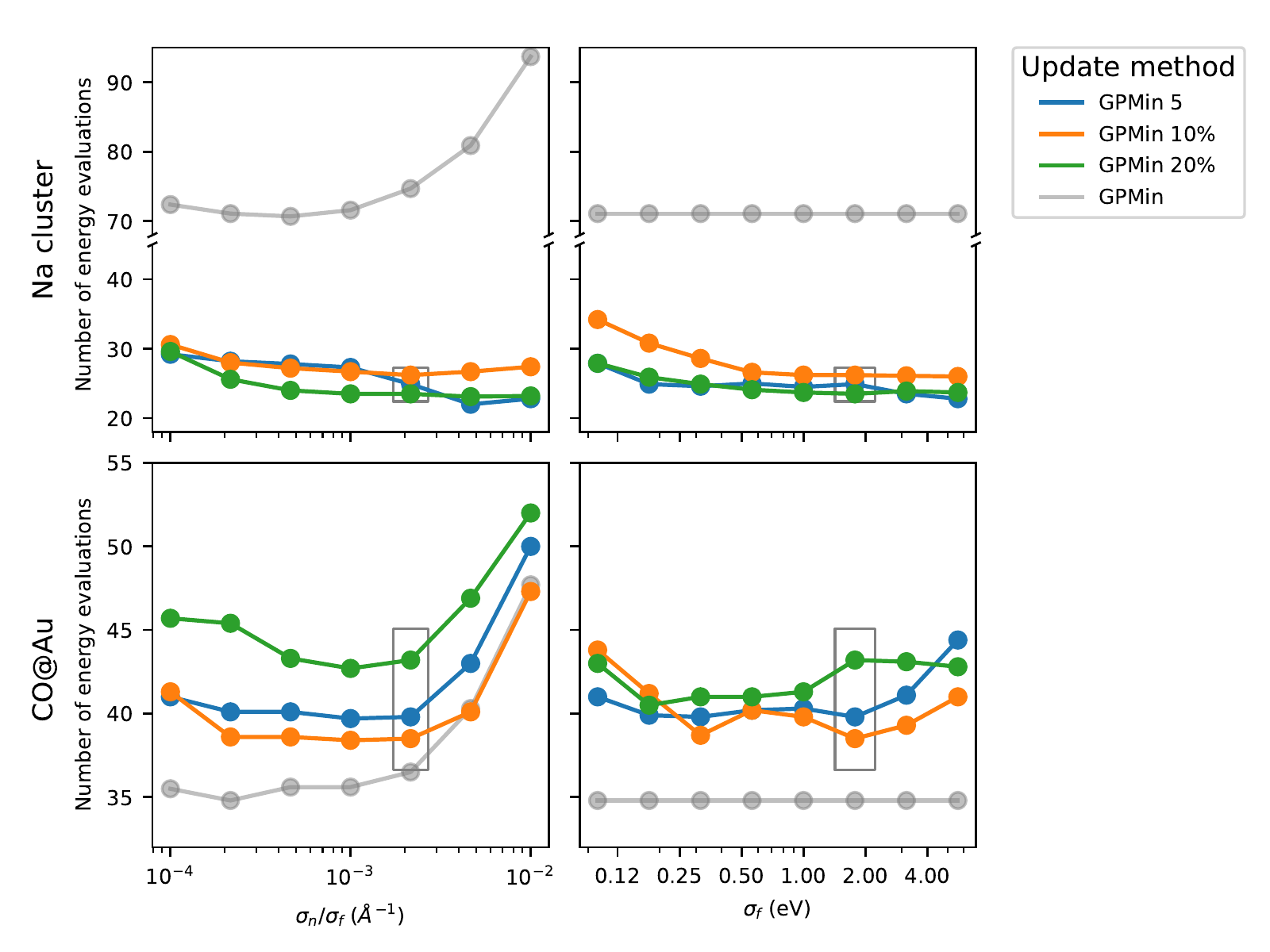}
    \caption{Average number of energy evaluations 
    needed to relax the two training set systems as a function of the 
    hyperparameter $\sigma_n/\sigma_f$ or of the initial value of $\sigma_f$,
    while the other one is kept fixed.
    The results are shown for the three different updating strategies
    and compared with the result of running  GPMin without update with the 
    same choice of hyperparameters. 
    The rectangles show the values of the hyperparameter that have been chosen 
    as default values. The value of $\sigma_f$ chosen in the right panels has been
    used in the relaxations shown in the left panels and, similarly, the value of $\sigma_n/\sigma_f$ that has been found optimal in the left panel is the one, which has been used in the relaxations in the right panel.}
    \label{fig:update_params}
\end{figure*} 
We now continue by considering the use of the GP optimizers more generally for systems with PESs described by DFT. Default values of the hyperparameters should be
chosen such that the algorithm performs well for a variety of atomic systems.  For
this purpose, we have chosen a training set consisting of two
different structures: (i) a 10-atom sodium cluster with random atomic
positions and (ii) a carbon dioxide molecule on a (111) surface with two
layers of gold and a $2\times2$ unit cell. We have generated 10 slightly
different initial configurations for each of the training systems by adding 
random numbers generated from a Gaussian distribution with standard deviation 
0.1 $\angstrom$. The training configurations are then relaxed using DFT energies and forces.

For each pair of the hyperparameters $(l,\sigma_n/\sigma_f)$, we relax the 
training systems and average over the number of DFT evaluations the optimizer
needs to find a local minimum. The results are shown in
Figure~\ref{fig:params}. The plot shows that the metallic cluster
benefits from relatively large scales, while the CO on gold system with tight CO bond requires
a shorter scale.  A too long
scale might even imply that the optimizer does not converge. The set
of hyperparameters $l=0.4 \,\angstrom$,
$\sigma_n = 1\ \textrm{meV}\angstrom^{-1}$ and $\sigma_f=1\ \textrm{eV}$ seems to be a good
compromise between the two cases and these are the default values we
shall use in the following.

A similar procedure has been used to determine the default values of the hyperparameters
and their initial values in the updated versions of GPMin. Here, the hyperparameter
$\sigma_n/\sigma_f$ is kept fixed during the optimization, whereas $l$ and $\sigma_f$ 
are determined using expression \eqref{eq:update}. The value of 
$\sigma_n/\sigma_f$ and the initial values of the other hyperparameters are then 
determined from the analysis of the performance of the optimizer on the two systems in 
the training set.

\begin{figure*}
    \centering
    \includegraphics{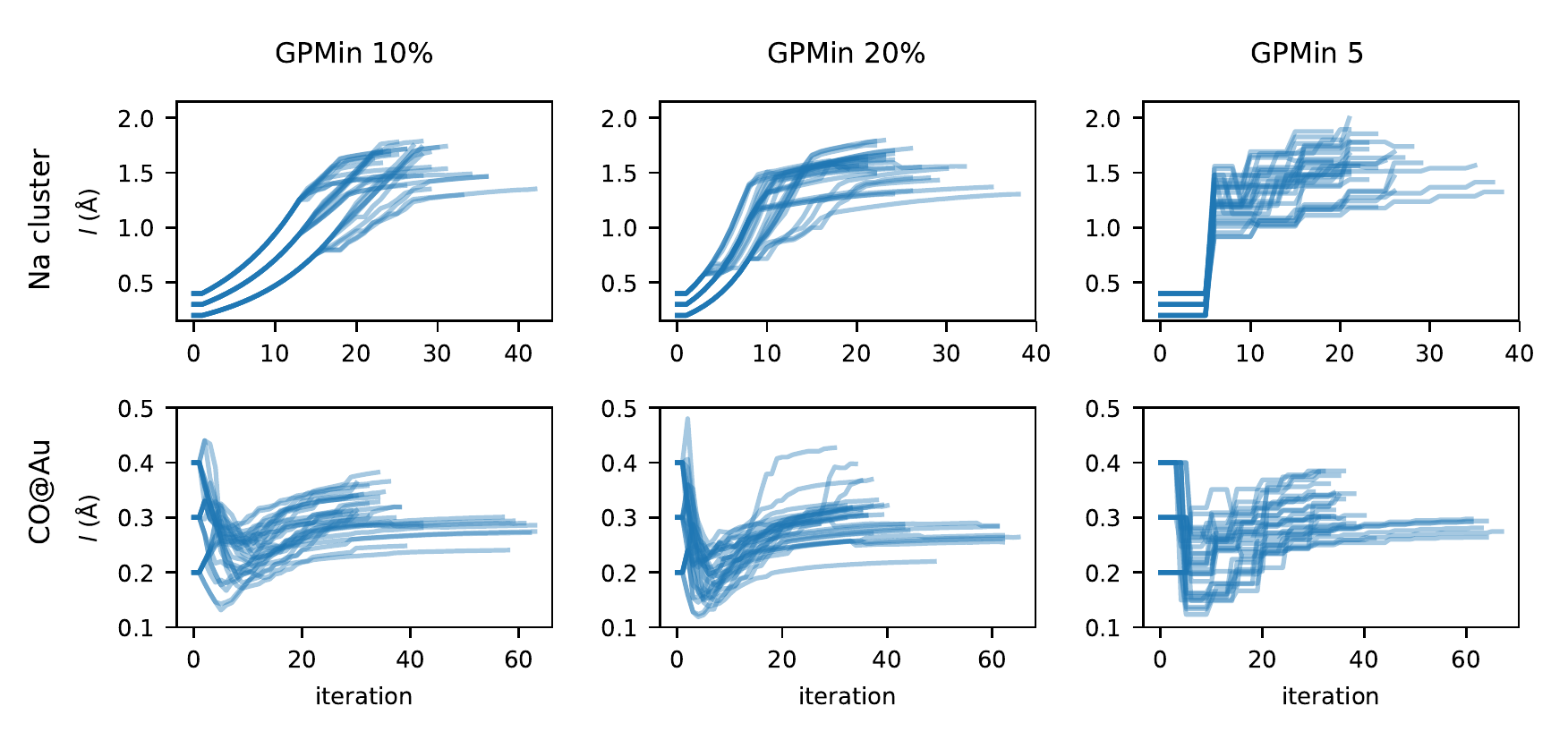}
    \caption{Evolution of the length scale $l$ with iteration for the three optimizers with update GPMin-5, GPMin-10\%, and GPMin-20\%.
    The upper panel shows the results for the sodium cluster, while the lower panel shows the evolution for the CO/Au system. In all
    cases three different values $l = 0.2, 0.3, 0.4 \angstrom$ for the initial scale has been considered. For the sodium cluster the length scale is seen to increase significantly, while in the case of the CO/Au system, the length scale first decreases and then subsequently increases. The final length scale various by about 30\% dependent on the particular initial structure of the systems.}
    \label{fig:scale}
\end{figure*}
The evolution of the hyperparameters depends on the details of the 
optimization of the marginal likelihood together with the frequency at which the 
hyperparameters are optimized. Here, we explore three different 
strategies: Unconstrained maximization of the marginal log-likelihood every 5 energy 
evaluations (``GPMin-5"), and two constrained optimization strategies, where the outcome of the
optimization is constrained to vary in the range $\pm 10\%$ and $\pm 20\%$ of the value
of the hyperparameter in the previous step (``GPMin-10\%" and ``GPMin-20\%", respectively). In the latter two cases we let the optimization take place 
whenever new information is added to the sample. The algorithm used to maximize
the marginal log-likelihood is L-BFGS-B \cite{lbfgsb} for all strategies.

We have relaxed the same 10 slightly different copies of the two training set systems 
described before using these three strategies for three different initial values of the 
scale (0.2, 0.3 and 0.4 $\angstrom$), 8 different initial values of $\sigma_f$ and
7 different values of the regularization parameter $\sigma_n/\sigma_f$. An overview of the full
results can be found in the Supplementary Material \cite{supplementary}.

The average numbers of energy evaluations needed to relax the training set for the different strategies and hyperparameters
are shown in Figure \ref{fig:update_params}. The initial value of the scale is chosen as $0.3\angstrom$. The plot shows the variation of the 
average number of energy evaluations with $\sigma_n/\sigma_f$ when the 
initial value of $\sigma_f=1.8 \textrm{eV}$ and the variation with $\sigma_f$ when the
value of $\sigma_n/\sigma_f=2 \times 10^{-3} \angstrom^{-1}$. The performance of the optimizers is seen to depend rather weakly on the parameter values in particular for the sodium cluster. We shall therefore in the following use the values $\sigma_f=1.8 \textrm{eV}$ and  $\sigma_n/\sigma_f=2 \times 10^{-3} \angstrom^{-1}$.

From the figure it can also be seen that the versions of the optimizer with updates perform considerably better than GPMin without updates for the sodium cluster, while for the CO molecule on gold, the version without update works slightly better than the three optimizers with updates.

To understand this behavior further we consider in Figure~\ref{fig:scale} the
evolution of the length scale $l$ as it is being updated. The scale is initially 
set at three different values $l = 0.2, 0.3, 0.4 \angstrom$. For the sodium 
cluster the update procedure quickly leads to a much longer length scale around 
$1.5 \angstrom$. For GPMin-5 the length scale is raised dramatically already at 
the first update after 5 energy evaluations, while for GPMin-10\% and GPMin-20\% 
the length scale increases gradually because of the constraint build into the 
methods. The advantage of a longer length scale is in agreement with the results 
above for the gold cluster described with the EMT interatomic interactions, where 
a long length scale also led to faster convergence. The situation is different 
for the CO/Au system, where the update leads first to a significant decrease in 
the scale and later to an increase saturating at a value around $0.3 \angstrom$. 
This result was to be expected
from the one shown in Figure \ref{fig:params} for the performance of GPMin without
hyperparameter update. We interpret the variation of the scale for the CO/Au system as being due to the different length scales present in the system, where the CO bond is short and strong while the metallic bonds are much longer. In the first part of the optimization the CO configuration is modified requiring a short scale, while the later stages involve the CO-metal and metal-metal distances. Overall the update of the scale does not provide an advantage over the GPMin without updates where the scale is kept fixed at $l = 0.4 \angstrom$. It can be seen that the final scales obtained, for example in the case of the sodium cluster optimized with GPMin-10\%, varies by about 30\%, where the variation depends on the particular system being optimized and not on the initial value for the length scale.

In the following we shall use $l = 0.3 \angstrom$ as the initial scale for the optimizers with updates. As shown in Figures S1, S2 and S3 in the supplementary material \cite{supplementary}, the results do not depend very much on the initial scale in the range $0.2 - 0.4 \angstrom$. Furthermore, the results for the EMT gold cluster indicate that long length scales should be avoided: it is easier for the algorithm to increase the length scale than to decrease it.

To summarize, we select the following default (initial) values of the hyperparameters for the updated versions of GPMin: $l=0.3\,\angstrom$, $\sigma_f = 2.0$ eV and $\sigma_n = 0.004 \,\textrm{eV}\angstrom^{-1}$ ($\sigma_n/\sigma_f = 0.002 \angstrom^{-1}$). These values are used in the rest of this paper.

\section{Results}
To test the Bayesian optimizers we have investigated their performance for seven different systems with DFT: a CO molecule on a Ag(111) surface, a C adsorbate on a Cu(100) surface, a distorted Cu(111) surface, bulk copper with random displacements of the atoms with Gaussian distribution and width 0.1 $\angstrom$, an aluminum cluster with 13 atoms in a configuration close to fcc; the $\rm{H}_2$ molecule, and the pentane molecule. All surfaces are represented by 2 layer slabs with a $2\times2$ unit cell and periodic boundary conditions along the slab. The bulk structure is represented by a $2\times2\times2$ supercell with periodic boundary conditions along the three unit cell vectors.  For each of the systems we have generated ten slightly different initial configurations by rattling the atoms by 0.1 \angstrom. The resulting configurations are then relaxed using the ASE and SciPy optimizers, together with the different GPMin optimizers.

It should be noted that in a few cases an optimizer fails to find a local minimum: an atomic configuration is suggested for which GPAW raises an error when it attempts to compute the energy, because two atoms are too close. This happens for SciPy's BFGS for one of the CO/Ag configurations and for SciPy's conjugate gradient method for one of the hydrogen molecule configurations.

\begin{figure*}
  \centering
  \includegraphics[scale=1]{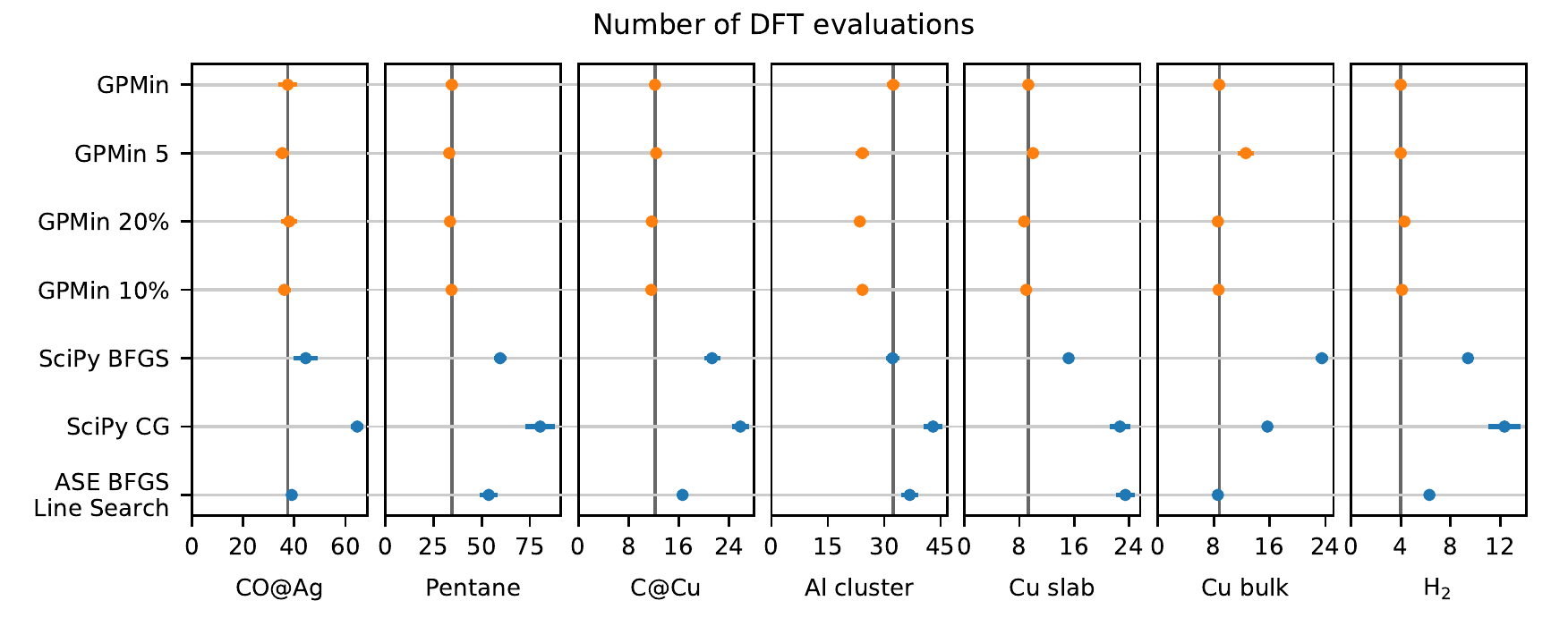}
  \caption{Number of DFT evaluations required to optimize a given
    structure. For each structure 10 different initial configurations
    are generated and optimized. The vertical line represents the 
    average number of steps of GPMin without parameter updates. The error bar represents the error on the average. A different color has been used to highlight the optimizers of the GPMin family.}
  \label{fig:lcao}
\end{figure*}

The results are collected in Figure~\ref{fig:lcao}. For the sake of clarity, ASE FIRE has been excluded of the plot, since it takes about a factor of three more steps than the fastest optimizer for all systems. The average number of DFT evaluations for the relaxation of the systems in the test set with the implementation of FIRE in ASE is $122 \pm 4$ for CO/Ag, $91\pm 5$ for the pentane molecule, $58\pm 4$ for C/Cu, $85\pm 3$ for the aluminum cluster, $62\pm 2$ for the Cu slab, $53\pm 1$ for Cu bulk and $30\pm 3$ for H${}_2$ molecule.

The GP optimizers are seen to compare favorably or on par with the best one of the other optimizers in all cases. GPMin without update is on average faster than the other optimizers for 6 of the 7 systems. 
For the bulk Cu system, it is only slightly slower than the ASE-BFGS algorithm. The updated GP optimizers perform even better with one exception: GPMin-5 is clearly worse than the other GP optimizers and ASE-BFGS for the copper bulk system. Since the atomic displacements from the perfect crystal structure are quite small ($\sim 0.1 \,\angstrom$), this system is probably within the harmonic regime and requires only few ($\sim 10$) iterations to converge. The ASE-BFGS can therefore be expected to perform well, which is also what is observed in Figure~\ref{fig:lcao}. GPMin-5 does not update the scale for the first 5 iterations, and when it does so, the new scale does not lead to immediate convergence. The plain GPMin and the two other optimizers with updates perform on par with ASE-BFGS.

Generally, the updated optimizers perform better than GPMin without updates, and both GPMin-10\% and GPMin-20\% with constrained update perform consistently very well. The updated optimizers are clearly better than the plain GPMin for the Al cluster similar to the behavior for the Na cluster used in the determination of hyperparameters. For the other training system, the CO/Au system, GPMin was seen to perform better than all the updated optimizers. However, in Figure~\ref{fig:update_params} the scale was chosen to be $l=0.3 \angstrom$, which is superior for that particular system. This behavior does not appear for any of the test systems including the CO/Ag system, which otherwise could be expected to be somewhat similar.

\section{Discussion}

We ascribe the overall good performance of the GP optimizers to their
ability to predict smooth potential energy surfaces covering both
harmonic and anharmonic regions of the energy landscape. Since the
Gaussian functions applied in the construction of the SPES
all have the scale $l$, the SPES will be harmonic at scales
much smaller than this around the minimum configuration. If the
initial configuration is in this regime the performance of the
optimizer can be expected to be comparable to BFGS, which is optimal
for a harmonic PES, and this is what is for example observed for the
Cu bulk system. We believe that the relatively worse performance of the SciPy implementation of BFGS can be attributed to an initial guess of the Hessian that is too far from the correct one.

Given the performance on both the training and test sets, GPMin-10\% seems to be a good choice. It should be noted that updating the hyperparameters require iteration over the the marginal log-likelihood leading to an increased computational cost. However, this is not a problem at least for systems comparable in size to the ones considered here.

The current version of the algorithm still has room for improvement. For example, different strategies for the update of hyperparameters may be introduced. Another, maybe even more interesting possibility, is to use more advanced prior models of the PES than just a constant. The prior model to the PES could for example be obtained from fast lower-quality methods. Somewhat along these lines there have been recent attempts to use previously known semi-empirical potentials for preconditioning more traditional gradient-based optimizers 
\cite{Tempkin-precon,gabor:2018preconditioners}. This approach might be combined with the GP framework suggested here.

We also note that the choice of the Gaussian kernel, even though encouraged by the characteristics of the resulting potential \cite{rasmussen2006a} and its previously reported success for similar problems \cite{hannesGP}, is to some extent arbitrary. It would be worthwhile to test its performance against other kernel functions, for example the Mat\'ern kernel, which has been reported to achieve better performance in different contexts \cite{Lizotte:2008:PBO:1626686, GP-CC,denzel2018gaussian}. The kernels used in the work here is also limited to considering only one length scale. More flexible kernels allowing for different length scales for different types of bonds would be interesting to explore.

The probabilistic aspect, including the uncertainty as expressed in Eq.~\eqref{eq:uncertainty}, is presently used only in the update of the hyperparameters. It could potentially lead to a further reduction of the number of function evaluations \cite{hannesGP}. The uncertainty
provides a measure of how much a region of configuration space has been explored and can thereby guide the search also in global optimization problems \cite{Lizotte:2008:PBO:1626686,deFreitas-review,Jorgensen:2018iu}.

Finally, a note on the limitations of the present version of the optimizer. The construction of the SPES involves the inversion of a matrix (Eq.~\ref{eq:prediction}) which is a square matrix, where the number of columns is equal to $n=N_c*(3*N+1)$, where $N$ is the number of atoms in the system and $N_c$ the number of previously visited configurations. This is not a problem for moderately sized systems, but for large systems, where the optimization also requires many steps, the matrix inversion can be very computationally time consuming, and the current version of the method will only be efficient if this time is still short compared to the time to perform the DFT calculations. In addition, this can also result in a memory issue for large systems where the relaxation takes many steps. These issues may be addressed by  considering only a subset of the data points or other sparsification techniques. Recently, Wang et al. \cite{Wang2019GPmillion} showed that by using the Blackbox Matrix-Matrix multiplication algorithm it is possible to reduce the cost of training from $O(n^3)$ to $O(n^2)$ and then by using distributed memory and 8 GPUs they were able to train a Gaussian process of $n\sim 4\times 10^4$ (this would correspond to about 100 steps for 150 atoms with no constraints) in 50 seconds. This time is negligible compared to the time for DFT calculations of systems of this size.

The GPMin optimizers are implemented in Python and available in ASE
\cite{ase}. 

\begin{acknowledgments}
We appreciate fruitful conversations with Peter Bj{\o}rn J{\o}rgensen.
This work was supported by Grant No. 9455 from VILLUM FONDEN. 
\end{acknowledgments}

\section{Appendix}
The optimization algorithm can be represented in pseudocode as follows:

\begin{algorithmic}
   \State\textbf{Input:} \\
Initial structure: $\mathbf{x}^{(0)} = (\mathbf{r}_1, \mathbf{r}_2, \dots, \mathbf{r}_N)$\\ 
Hyperparameters: $l$, $\sigma_n$, \\
Tolerance: $f_\text{max}$\\
   \State $E^{(0)}, \mathbf{f}^{(0)} \gets $\Call{Calculator}{$\mathbf{x}^{(0)}$}
   \State $E_p \gets E^{(0)}$
   \While{$\max_{i}\vert \mathbf{f}_i^{(0)}\vert> f_\text{max}$}
      \State $X, Y \gets$ \Call{Update}{$\mathbf{x}^{(0)}, E^{(0)}, \mathbf{f}^{(0)}$}
      \State $E_p \gets \max Y_E$ 
      \State $\mathbf{x}^{(1)} \gets $ \Call{l-bfgs-b}{GP($X,Y$), start\_from = $\mathbf{x}^{(0)}$}
      \State $E^{(1)}, \mathbf{f}^{(1)} \gets $\Call{Calculator}{$\mathbf{x}^{(1)}$}
      \While{$E^{(1)} > E^{(0)}$}
         \State $X, Y \gets$ \Call{Update}{$ \mathbf{x}^{(1)}, E^{(1)}, \mathbf{f}^{(1)}$}
         \State $E_p \gets \max Y_E$
         \State $\mathbf{x}^{(1)} \gets $ \Call{l-bfgs-b}{GP($X,Y$), start\_from = $\mathbf{x}^{(0)}$}
         \State $E^{(1)}, \mathbf{f}^{(1)} \gets $\Call{Calculator}{$\mathbf{x}^{(1)}$}
         \If{$\max_{i}\vert \mathbf{f}_i^{(1)}\vert> f_\text{max}$}  break \EndIf
      \EndWhile
      \State $\mathbf{x}^{(0)}, E^{(0)}, \mathbf{f}^{(0)} \gets \mathbf{x}^{(1)}, E^{(1)}, \mathbf{f}^{(1)} $
   \EndWhile
\State\textbf{Output:} $\mathbf{x}^{(0)}, E^{(0)}$
\end{algorithmic}

\bibliography{references}


\end{document}